# Sub-Alfvénic Solar Wind observed by PSP: Characterization of Turbulence, Anisotropy, Intermittency, and Switchback

R. Bandyopadhyay[1], W. H. Matthaeus[2,3], D. J. McComas[1], R. Chhiber[2,4], A. V. Usmanov[2,4], J. Huang[5], R. Livi[6], D. E. Larson[6], J. C. Kasper[5,7], A. W. Case[8], M. Stevens[8], P. Whittlesey[6], O. M. Romeo[6,9], S. D. Bale[6,9,10,11], J. W. Bonnell[6], T. Dudok de Wit[12], K. Goetz[13], P. R. Harvey[6], R. J. MacDowall[4], D. M. Malaspina[14], M. Pulupa[6]

[1]Department of Astrophysical Sciences, Princeton University, Princeton, NJ 08544, USA
[2]Department of Physics & Astronomy, University of Delaware, DE 19716, USA
[3]Bartol Research Institute, University of Delaware, Newark, DE 19716, USA
[4]NASA Goddard Space Flight Center, Greenbelt, MD 20771, USA
[5]Climate and Space Sciences and Engineering, University of Michigan, Ann Arbor, MI 48109, USA
[6]Space Sciences Laboratory, University of California, Berkeley, CA 94720-7450, USA
[7]BWX Technologies, Inc., Washington DC 20002, USA
[8]Smithsonian Astrophysical Observatory, Cambridge, MA 02138, USA
[9]Physics Department, University of California, Berkeley, CA 94720-7300, USA
[10]The Blackett Laboratory, Imperial College London, London, SW7 2AZ, UK
[11]School of Physics and Astronomy, Queen Mary University of London, London E1 4NS, UK
[12]LPC2E, CNRS and University of Orléans, Orléans, France
[13]School of Physics and Astronomy, University of Minnesota, Minneapolis, MN 55455, USA
[14]Laboratory for Atmospheric and Space Physics, University of Colorado, Boulder, CO 80303, USA

**Abstract**

In the lower solar coronal regions where the magnetic field is dominant, the Alfvén speed is much higher than the wind speed. In contrast, the near-Earth solar wind is strongly super-Alfvénic, i.e., the wind speed greatly exceeds the Alfvén speed. The transition between these regimes is classically described as the "Alfvén point" but may in fact occur in a distributed Alfvén critical region. NASA's Parker Solar Probe (PSP) mission has entered this region, as it follows a series of orbits that gradually approach more closely to the sun. During its 8th and 9th solar encounters, at a distance of $\approx 16\,R_\odot$ from the Sun, PSP sampled four extended periods in which the solar wind speed was measured to be smaller than the local Alfvén speed. These are the first in-situ detections of sub-Alfvénic solar wind in the inner heliosphere by PSP. Here we explore properties of these samples of sub-Alfvénic solar wind, which may provide important previews of the physical processes operating at lower altitude. Specifically, we characterize the turbulence, anisotropy, intermittency, and directional switchback properties of these sub-Alfvénic winds and contrast these with the neighboring super-Alfvénic periods.

## 1. Introduction

The solar wind consists of highly ionized, magnetized plasma that flows from the Sun's corona to the interplanetary space (Parker 1958). The birthplace of the wind, deep in the corona, is a magnetically dominated region, as is evident in coronagraph images (Cranmer & Winebarger 2019). A manifestation of this control is that the Alfvén speed, the speed at which the dominant magnetohydrodynamic (MHD) signals propagate in the magnetized coronal plasma, is greater than other dynamically important speeds such as wind speed. However, as the solar plasma expands from the corona into space, the magnetic field cedes control of the coronal plasma, so that additional in-situ dynamical processes, involving plasma flows and turbulence, become relatively more important. This transition is traditionally viewed as occurring where the speed of the wind, undergoing continual acceleration, exceeds the Alfvén speed. For a variety of reasons, plasma properties, and in particular properties of fluctuations, may differ above and below, this transition region (DeForest et al. 2016; Ruffolo et al. 2020; see Fig 2 of Fox et al. 2016). As the Parker Solar Probe (PSP) mission (Fox et al. 2016; McComas et al. 2007) descends into the deep corona, it becomes possible to observe the plasma properties across this transition region.

## 2. Background

A preponderance of coronal fluctuations are believed to be in low-compressibility Alfvén modes (Belcher & Davis Jr. 1971). When the wind speed exceeds the Alfvén speed, these modes can no longer communicate information to lower altitudes. There are no reliable determinations of the magnetic field strength throughout most of the lower corona. Therefore, even though densities are relatively well known in the coronal regions, Alfvén speeds are not. In the lower corona the magnetic field is expected to be dominant over the kinetic energy so that the Alfvén speed ($V_A$) is expected to be much higher than the wind speed. Eventually, the solar wind speed ($V_{sw}$) becomes higher than the Alfvén speed, and near 1 au we typically observe $V_{sw}/V_A \sim 10$ (Borovsky et al. 2019; Klein & Vech 2019). The Alfvén critical points refer to the idealized case for which fluctuations are absent and the smoothly varying Alfvén and flow speeds become equal to one another (e.g., Chhiber et al. 2019; Kasper & Klein 2019). In more realistic inhomogeneous and nonstationary conditions (Chhiber et al. 2021; DeForest et al. 2016, 2018),

Following recent reports (Kasper et al. 2021) that this "boundary" between sub-Alfvénic and super-Alfvénic flow has been crossed on several occasions, here we analyze the turbulence properties within these special sub-Alfvénic regions, and contrast these with properties of nearby standard super-Alfvénic solar wind.

the Alfvénic transition is expected to occur irregularly over a more extended region.

NASA's PSP Mission was launched in August 2018 to reach closer to the Sun than any other spacecraft before. With the launch of PSP, the Alfvénic critical point has been a topic of increasingly frequent discussion in coronal and solar wind physics. The nature of the plasma and its fluctuations in this region will reveal important information about the fluctuations and waves emanating from lower altitudes.

During the $8^{th}$ orbit around the Sun, near the perihelion, PSP sampled three extended periods of solar wind where the wind speed became smaller than the local Alfvén speed. Also, during the $9^{th}$ solar encounter, PSP observed another extended sub-Alfvénic period near the perihelion. Although there had been a few detections of sub-Alfvénic wind near the Earth (Gosling et al. 1982; Smith et al. 2001; Stansby 2021, 20; Usmanov et al. 2005), these measurements by PSP provide the first opportunity to study and characterize the lower coronal solar wind plasma.

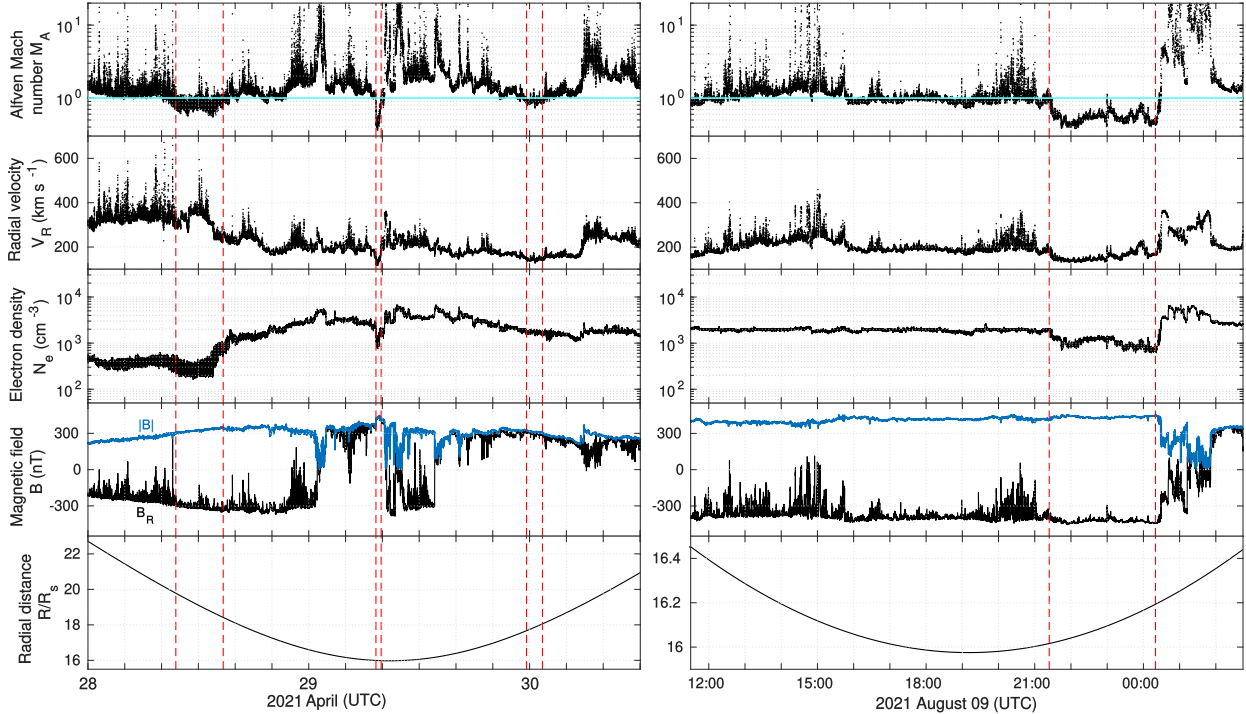

**Figure 1.** An overview of plasma parameters surrounding the sub-Alfvénic intervals observed by PSP near its 8[th] (left) and 9[th] (right) perihelion. The panels from top show Alfvén Mach number $M_A = V_R/V_A$ where $V_A$ is the Alfvén speed (see text), radial component of solar wind proton velocity $V_R$, electron number density $N_e$, radial component and magnitude of magnetic field $\boldsymbol{B}$, and heliocentric distance in units of solar radii. A cyan line is drawn at $M_A = 1$ in the top panels for visual ease. The dashed red lines indicate the sub-Alfvénic periods.

### 3. PSP Data During Encounters 8 & 9

We use magnetic-field ($\boldsymbol{B}$) data from the flux-gate magnetometer (MAG) from the FIELDS instrument suite (Bale et al. 2016, 2019, 2020). Proton radial velocity ($V_R$) are obtained from partial moments from the Solar Probe ANalyzer for Ions (SPAN-I) on the SWEAP instrument suite (Kasper et al. 2016, 2019; Livi et al. 2021, 2020). FIELDS electron density ($N_e$) are derived from the quasi-thermal noise (QTN) spectrum measured by the Radio Frequency Spectrometer onboard PSP (Moncuquet et al. 2020). The local Alfvén speed is computed as $V_A = |\boldsymbol{B}|/\sqrt{\mu_0 m_p N_e}$, where $\mu_0$ is the magnetic permeability of vacuum and $m_p$ is the mass of proton. The local Alfvén Mach number is $M_A = V_R/V_A$.

The relevant plasma variables, covering a period of about 3 days centered on the 8th perihelion around the Sun, are shown in the left panel of Fig. 1. The dashed red lines indicate the three periods when $M_A<1$ for majority of the time. The times of sub-Alfvénic periods are selected as 2021 April 28 09:33 to 14:42 UTC, 2021 April 29 07:18 to 07:52 UTC, and 2021 April 29 23:40 to 2021 April 30 01:24 UTC. A detailed description of these intervals has been presented in (Kasper et al. 2021). In addition, there is a sub-Alfvénic interval near the 9[th] perihelion (right panels of Fig.1), approximately from 2021 August 09 21:24 to 2021 August 10 00:20 UTC. In the following sections, we compare plasma properties of the four sub-Alfvénic periods and compare them with the neighboring super-Alfvénic periods shown in Fig. 1. The super-Alfvénic intervals from encounter 8 are selected from 2021 April 28 00:00 to 09:33 UTC; 2021 April 28 4:42 to 2021 April 29 07:18 UTC; 2021 April 29 07:52 to 23:40; and 2021 April 30 01:24 to 12:00 UTC. For encounter 9, we select the super-Alfvénic periods as 2021 August 09 11:30 to 21:24 and 2021 August 10 00:20 to 02:45 UTC. We present statistics of the turbulence amplitude, variance anisotropy, intermittency, and switchback properties in the two kinds of samples.

## 4. Methodology & Results
### 4.1 Turbulence

We begin with the turbulence amplitude of the fluctuations. We use the magnetic field to measure the turbulence amplitude as

$$\delta B = \sqrt{\langle |\boldsymbol{B}(t) - \langle \boldsymbol{B}\rangle|^2\rangle},$$

where $\langle \cdots \rangle$ is a time average, over some chosen time range, usually several correlation times. Here we choose $\approx 10$ min intervals and evaluate the turbulence amplitude in each interval. Then, we accumulate the intervals separately for sub-Alfvénic or super-Alfvénic conditions (Fig. 1). Fig. 2 shows the histograms of frequency of occurrence of the turbulence level (in logarithm), separated into these two categories. The $\delta B$ is in the usual nT units, and the logarithm is base 10.

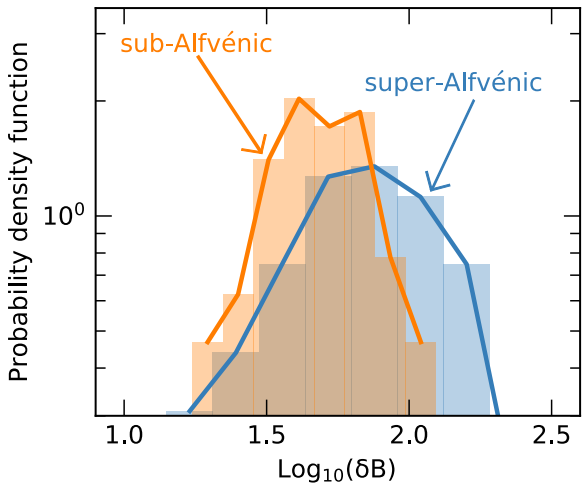

**Figure 2.** Probability distribution function (PDF) of magnetic field turbulence amplitude in the sub-Alfvénic solar wind and the neighboring super-Alfvénic solar wind intervals observed by PSP (see Fig. 1 and the text in Sec. 3).

From Fig. 2, there is a clear difference in the turbulence amplitude in the sub and super Alfvénic wind samples. The super-Alfvénic samples are considerably more turbulent. The most probable value for the combined sub-Alfvenic samples is $\delta \tilde{B} = 52.5$ nT; a larger value of $\delta \tilde{B} = 75.5$ nT is found to be most probable for the combined super-Alfvénic intervals. Likewise, the average values are $\delta \bar{B} = 52.4$ nT with standard deviation of $\sigma_{\delta B}=23.2$ nT for the sub-Alfvénic samples but has a larger value of $\delta \bar{B} = 79.9$ nT with standard deviation of $\sigma_{\delta B}=48.6$ nT for the super-Alfvénic intervals. While it is unclear yet if this decrease in turbulence level is typical of the sub-Alfvénic lower corona, a Kolmogorov-Smirnov (KS) two-sample test yields a p-value less than 0.001. The very small p-value obtained from the KS test suggests that it is highly unlikely that the sub- and super-Alfvénic observations are drawn from the same underlying distribution. This result suggests that the super-Alfvénic wind may be less magnetically constrained, thus indicating a tendency towards larger amplitude fluctuations.

### 4.2 Variance Anisotropy

The interplanetary magnetic field variance anisotropy (Oughton et al. 2016; Parashar et al. 2016) measures the departure from equal magnetic field component energies $\langle b_x^2\rangle = \langle b_y^2\rangle = \langle b_z^2\rangle$. Note that the mean magnetic field is excluded, usually by subtracting the average value $\boldsymbol{B}_0$. Variance anisotropy is typically defined in a particular Cartesian coordinate system, as

$$A_b = \frac{\langle b_x^2 + b_y^2\rangle}{\langle b_\parallel^2\rangle},$$

where $\boldsymbol{B}_0$ is along the z-axis and the magnetic fluctuations are $\boldsymbol{b} = (b_x, b_y, b_\parallel)$. A value of $A_b = 2$ corresponds to isotropic distribution.

Values of $A_b > 2$ denote transverse anisotropy of the type associated with Alfvén modes at small amplitude (Barnes 1979). Parallel variance (at small amplitude) indicates the presence of magnetosonic modes and introduces the possibility of compressive motions. So, variance anisotropy is often associated with "magnetic compressibility."

To proceed, we divide the PSP data in $\approx 10$ min intervals and compute the variance anisotropy in each sample. A quantitative measure of variance anisotropy occurrence is compared for sub-Alfvénic periods and a sample of super-Alfvénic period in Fig. 3. The Figure shows a binned histogram (frequency of occurrence) of log($A_b$).

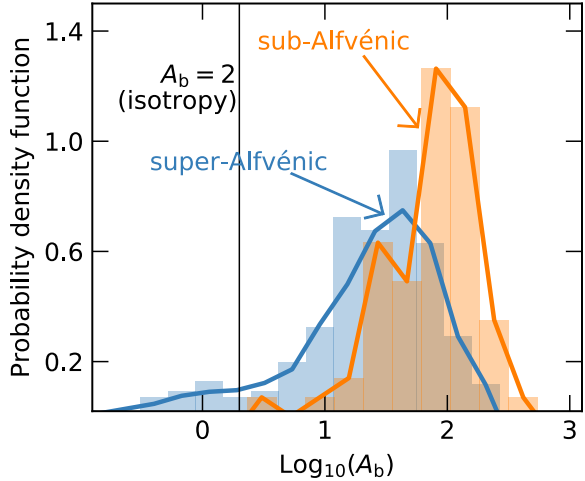

**Figure 3.** Distribution of variance anisotropy in sub-Alfvénic and the neighboring super-Alfvénic solar wind intervals observed by PSP near perihelion 8 and 9 (see Fig. 1 and the text in Sec. 3). The vertical line indicates the value $A_b = 2$, which corresponds to isotropic distribution.

It is evident that both regions are highly anisotropic ($A_b \gg 2$), but the sub-Alfvénic solar wind samples are systematically more anisotropic than the neighboring super-Alfvénic regions of the solar wind. The peak of the distribution is $\widetilde{A_b}$ = 43.2 for the super-Alfvénic periods but has a larger most probable anisotropy $\widetilde{A_b}$ = 139.7 is found for the sub-Alfvénic distribution. Similarly, the mean values are $\overline{A_b}$ = 39.9 with standard deviation of $\sigma_{A_b}$=40.7 for the super-Alfvénic periods and $\overline{A_b}$ = 98.3 with standard deviation of $\sigma_{A_b}$=82.2 for the sub-Alfvénic samples. A Kolmogorov-Smirnov test gives p-value less than 0.001, supporting the difference of the two population.

We note that a large variance anisotropy is a common expectation in models of coronal heating in open field line regions (e.g., Einaudi et al. 1996; also see Habbal et al. 1995; Cranmer 2018) for which heating is a result of turbulence and current sheet formation induced by launching of transverse fluctuation due to photospheric deflection of field lines.

### 4.3 *Intermittency*

Intermittency refers to occurrence of extreme events distributed in a super-Gaussian manner. Familiar signatures of intermittency are revealed by the Partial Variance of Increment (PVI) method, a widely used measure to detect the occurrence of sharp gradients in the turbulent fields (such as magnetic field) (Greco et al. 2012, 2017). Examples include current sheets and sites of magnetic reconnection. These regions of sharp gradients, idealized as discontinuities, are found to be sites of enhanced dissipation, particle heating (e.g., Bandyopadhyay et al. 2020; Chasapis et al. 2015) and acceleration (e.g., Bandyopadhyay et al. 2020b; Tessein et al. 2013) in space plasmas.

The PVI of the magnetic field $\boldsymbol{B}$ at time t is defined, for time lag τ, as (Greco et al. 2017):

$$\text{PVI}_{t,\tau} = \frac{|\Delta \boldsymbol{B}(t,\tau)|}{\sqrt{\langle|\Delta \boldsymbol{B}(t,\tau)|^2\rangle}},$$

The temporal increment of the magnetic field is defined as $\Delta \boldsymbol{B}(t,\tau) = \boldsymbol{B}(t+\tau) - \boldsymbol{B}(t)$.

The PVI is similar to the first-order structure function, but is distinct in that it is a pointwise measure instead of an averaged measure. Values of PVI > 2.5 are associated with non-Gaussian structures, such as current sheets and reconnection sites (Greco et al. 2017). Progressively higher PVI values are less likely to be random events drawn from a Gaussian distribution.

To compute the PVI, we use a moving average of 10 min, and for the increment, we use a lag of τ = 1 s. PVI values collected from the sub-Alfvénic and the neighboring super-Alfvénic solar wind intervals display slightly different statistical distributions, as illustrated in Fig. 4.

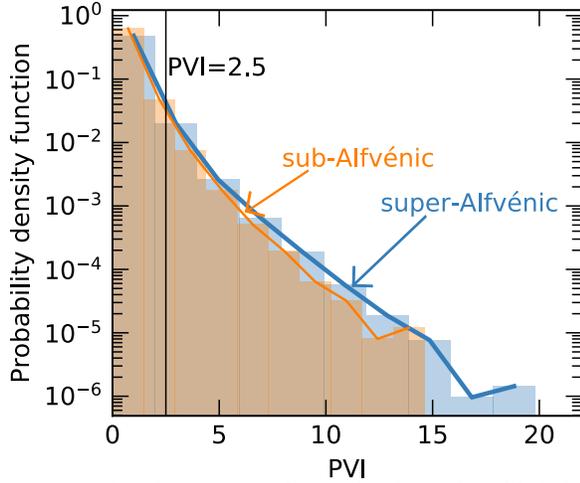

**Figure 4.** Histogram of PVI in the sub-Alfvénic and the neighboring super-Alfvénic periods near perihelion 8 and 9 (see Fig. 1 and the text in Sec. 3). PVI values larger than 2.5 (vertical line) represent intermittent structures.

The figure directly affects the comparison by superposing the normalized histograms of PVI occurrence rates in the respective samples. Unlike turbulence level and variance anisotropy, PVI appears to indicate only a slight difference in the two kinds of samples. We do note a small excess of large values present in the super-Alfvénic distribution, and a Kolmogorov-Smirnov test does give p-value < 0.001 supporting that the two samples are drawn from different distribution.

### 4.4 Switchbacks

Switchbacks have been a topic of increasing interest in PSP observations. Switchbacks are sudden large polarity reversals in magnetic field accompanied by simultaneous increase in velocity. Switchbacks were observed by spacecraft before PSP (e.g., Matteini et al. 2015; Neugebauer & Goldstein 2013), but in the initial PSP orbits they are abundantly observed near the encounter phase (Bale et al. 2019; Dudok de Wit et al. 2020). Switchbacks may be important in understanding coronal heating (Hernández et al. 2021), solar wind expansion and acceleration, as well as energetic particle acceleration (Bandyopadhyay et al. 2021).

Magnetic-field deflections, including switchbacks, may be quantified following (Dudok de Wit et al. 2020), by the parameter

$$z = \frac{1}{2}(1 - \cos(\alpha)),$$

where $\cos(\alpha) = \mathbf{B} \cdot \langle \mathbf{B} \rangle / |\mathbf{B}||\langle \mathbf{B} \rangle|$ and the brackets denote as suitable local or regional average. Here, we use an averaging interval of 10 min. The Z variable can admit values between 0 and 1. Values of $Z > 1/2$ indicate that the field is in a polarity-reversed state (i.e., a switchback) and lower values correspond to "background" magnetic polarity.

Fig. 5 illustrates the distribution of Z values (in logarithm) for the sub-Alfvénic and super-Alfvénic periods. Fig. 5 shows that magnetic deflections, measured by the Z parameter, are smaller in the sub-Alfvénic periods than in the super-Alfvénic parts.

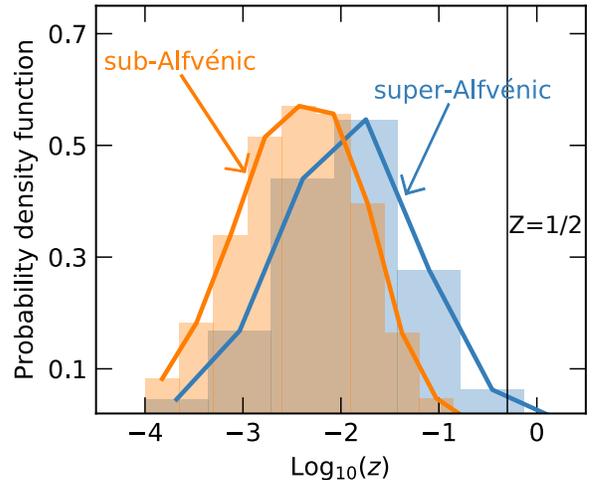

**Figure 5.** Histogram of switchback parameter (Z) in the sub-Alfvenic solar wind and the neighboring super-Alfvénic solar wind intervals observed by PSP (see Fig. 1 and the text in Sec. 3). The vertical line indicates the value Z = ½, which correspond to polarity-reversed magnetic field. Higher Z values represent stronger switchback.

Similar to turbulence amplitude (Sec. 4.1) and variance anisotropy (Sec. 4.2) the most probable values of the deflection parameter Z are different. No samples with $Z > 1/2$ occur in the sub-Alfvénic samples. The peak of the distribution is $\tilde{Z} = 0.0084$ for the sub-Alfvénic periods and $\tilde{Z} = 0.018$ for the super-Alfvénic periods. Further, the mean values are $\bar{Z} = 0.051$ with a standard deviation of $\sigma_Z = 0.0039$ for the super-Alfvénic periods and $\bar{Z}$ = 0.0096 with a standard deviation of $\sigma_z$=0.011 for the sub-Alfvénic periods. A Kolmogorov-Smirnov test yields p-value < 0.001 indicating the difference in the two distributions. This result indicates that there are stronger switchbacks in the super-Alfvénic solar wind, as predicted by some of the models proposed to explain the origins of switchbacks (Ruffolo et al. 2020; Schwadron & McComas 2021).

## 5. Discussion and Conclusions

PSP is a historic and unique mission by NASA. A main science objective of the mission is to "determine the structure and dynamics of the plasma and magnetic fields at the sources of the solar wind." The results presented in this paper advance key steps towards that goal. In this regard, a distinction worth considering is that of the prevailing physics in sub-Alfvénic versus the super-Alfvénic regions. It is tempting to view the (presumably) low beta, rigidly rotating (Weber & Davis 1967), magnetically dominated plasma closer to the sun as the true coronal plasma. Then the more distant and much better explored higher beta, flow-dominated, non-rigidly rotating plasma is the solar wind plasma. Among many other anticipated discoveries, PSP is set to explore the differences in these regions (McComas et al. 2007). There is a widespread anticipation that observation of the magnetically dominated (lower) corona will provide the clearest clues yet concerning the processes that heat the coronal plasma, accelerate the wind, and affect transport of energetic particles (McComas et al. 2016, 2019). The present paper represents an early effort to distinguish fluctuation properties in these two regions, taking advantage of the first glimpses afforded by PSP.

In Table 1 we recapitulate some of the statistical comparisons between the two regions that were considered in the previous sections. There are prominent distinctions between super- and sub-Alfvénic regions for all but one of the quantities (i.e., PVI) we examined. Indeed, the differences observed appear to fit nicely into the conventional views concerning "coronal" vs "solar wind" plasma.

**Table 1.** Average values of the accumulated sub-Alfvénic and super-Alfvénic solar wind samples.

|  | Sub-Alfvénic | Super-Alfvénic |
|---|---|---|
| Turbulence amplitude $\overline{\delta B}$ | 52.4 nT | 79.9 nT |
| Variance anisotropy $\overline{A_b}$ | 98.3 | 39.9 |
| $\overline{PVI}$ | 0.65 | 0.66 |
| Switchback parameter $\bar{Z}$ | 0.0096 | 0.051 |

We note here that the turbulence amplitude, variance anisotropy measure, and PVI all are sensitive to the size of the averaging interval (e.g., Isaacs et al. 2015), and thus the different sample size of the sub- and super- Alfvénic intervals may affect the statistical results. For example, we have included substantially more super-Alfvenic data. However, as pointed out by (Kasper et al. 2021), the sub-Alfvénic intervals are all of greater duration than turbulence correlation time, suggesting that the statistical characterizations are at least moderately stable. As a test, we repeated our analyses using the same number of minutes of both kinds of solar wind. For each sub-Alfvénic interval, we choose half of the preceding and half of the following super-Alfvenic interval. We found that our results do not change by this procedure. This finding supports the robustness of the results. Further, for each characterized quantity, we find that the difference between the averages of the super-Alfvenic and all of the sub-Alfvenic intervals is larger than the standard deviation of the averages of the averages calculated using the individual sub-Alfvénic intervals. Therefore, we think that it is reasonable to combine the sub-Alfvénic intervals into a single ensemble when calculating statistical properties.

Future PSP observations along with global solar wind models will shed more light on the nature of the transition from sub to super- Alfvénic flow, and whether it is best described as a wrinkled surface or an extended and fragmented zone (Chhiber et al. 2022).

## Acknowledgments
We are deeply indebted to everyone that helped make the Parker Solar Probe (PSP) mission possible. This work was supported as a part of the PSP mission under contract NNN06AA01C. This research was partially supported by the Parker Solar Probe Plus project through Princeton/IS⊙IS subcontract SUB0000165, in part by PSP GI grant 80NSSC21K1765 at the University of Delaware, and in part by PSP GI grant 80NSSC21K1767 at Princeton University. Parker Solar Probe was designed, built, and is now operated by the Johns Hopkins Applied Physics Laboratory as part of NASA's Living with a Star (LWS) program (contract NNN06AA01C). Support from the LWS management and technical team has played a critical role in the success of the Parker Solar Probe mission.

## References

Bale, S., Badman, S., Bonnell, J., et al. 2019, Nature (Nature Publishing Group), 1

Bale, S. D., Goetz, K., Harvey, P. R., et al. 2016, Space Sci Rev, 204, 49

Bale, S. D., MacDowal, R. J., Koval, A., et al. 2020, https://doi.org/10.48322/wqcs-a534

Bandyopadhyay, R., Matthaeus, W. H., McComas, D. J., et al. 2021, Astroomy Astrophys, 650, L4

Bandyopadhyay, R., Matthaeus, W. H., Parashar, T. N., et al. 2020a, Phys Rev Lett, 124 (American Physical Society), 255101

Bandyopadhyay, R., Matthaeus, W. H., Parashar, T. N., et al. 2020b, Astrophys J Suppl Ser, 246 (American Astronomical Society), 61

Barnes, A. 1979, in Solar System Plasma Physics, vol. I, ed. E. N. Parker, C. F. Kennel, & L. J. Lanzerotti (Amsterdam: North-Holland), 251

Belcher, J. W., & Davis Jr., L. 1971, J Geophys Res 1896-1977, 76, 3534

Borovsky, J. E., Denton, M. H., & Smith, C. W. 2019, J Geophys Res Space Phys, 124, 2406

Chasapis, A., Retino, A., Sahraoui, F., et al. 2015, Astrophys J Lett, 804, L1

Chhiber, R., Matthaeus, W. H., Usmonov, A., et al. 2022, MNRAS (in prep.)

Chhiber, R., Usmanov, A. V., Matthaeus, W. H., & Goldstein, M. L. 2019, Astrophys J Suppl Ser, 241 (American Astronomical Society), 11

Cranmer, S. R. 2018, Res Notes AAS, 2 (American Astronomical Society), 158

Cranmer, S. R., & Winebarger, A. R. 2019, Annu Rev Astron Astrophys, 57, 157

DeForest, C. E., Howard, R. A., Velli, M., Viall, N., & Vourlidas, A. 2018, \apj, 862, 18

DeForest, C. E., Matthaeus, W. H., Viall, N. M., & Cranmer, S. R. 2016, Astrophys J, 828 (American Astronomical Society), 66

Fox, N. J., Velli, M. C., Bale, S. D., et al. 2016, Space Sci Rev, 204, 7

G. Einaudi, M. Velli, H. Politano, & Pouquet, A. 1996, Astrophys J, 457

(American Astronomical Society), https://doi.org/10.1086/309893

Gosling, J. T., Asbridge, J. R., Bame, S. J., et al. 1982, J Geophys Res Space Phys, 87, 239

Greco, A., Matthaeus, W. H., Perri, S., et al. 2017, Space Sci Rev, 214, 1

Greco, A., Valentini, F., Servidio, S., & Matthaeus, W. H. 2012, Phys Rev E, 86 (American Physical Society), 066405

Hernández, C. S., Sorriso-Valvo, L., Bandyopadhyay, R., et al. 2021, Astrophys J Lett, 922 (American Astronomical Society), L11

Isaacs, J. J., Tessein, J. A., & Matthaeus, W. H. 2015, J Geophys Res Space Phys, 120, 868

Kasper, J., Bale, S., Belcher, J., et al. 2019, Nature (Nature Publishing Group), 1

Kasper, J. C., Abiad, R., Austin, G., et al. 2016, Space Sci Rev, 204, 131

Kasper, J. C., & Klein, K. G. 2019, Astrophys J Lett, 877 (American Astronomical Society), L35

Kasper, J. C., Klein, K. G., Lichko, E., et al. 2021, Phys Rev Lett, 127 (American Physical Society), 255101

Klein, K. G., & Vech, D. 2019, Res Notes AAS, 3 (American Astronomical Society), 107

Livi, R., Larson, D. E., Kasper, J. C., et al. 2021, Earth Space Sci Open Arch, 20

Livi, R., Larson, D. E., & Rahmati, A. 2020, https://doi.org/10.48322/ypyh-s325

Matteini, L., Horbury, T. S., Pantellini, F., Velli, M., & Schwartz, S. J. 2015, Astrophys J, 802 (American Astronomical Society), 11

McComas, D., Christian, E., Cohen, C., et al. 2019, Nature (Nature Publishing Group), 1

McComas, D. J., Alexander, N., Angold, N., et al. 2016, Space Sci Rev, 204, 187

McComas, D. J., Velli, M., Lewis, W. S., et al. 2007, Rev Geophys, 45, https://agupubs.onlinelibrary.wiley.com/doi/abs/10.1029/2006RG000195

Moncuquet, M., Meyer-Vernet, N., Issautier, K., et al. 2020, Astrophys J Suppl Ser, 246 (American Astronomical Society), 44

Neugebauer, M., & Goldstein, B. E. 2013, AIP Conf Proc, 1539, 46

Oughton, S., Matthaeus, W. H., Wan, M., & Parashar, T. 2016, J Geophys Res Space Phys, 121, 5041

Parashar, T. N., Oughton, S., Matthaeus, W. H., & Wan, M. 2016, Astrophys J, 824 (American Astronomical Society), 44

Parker, E. N. 1958, Astrophys J, 128, 664
Rifai Habbal, S., Esser, R., Guhathakurta, M., & Fisher, R. R. 1995, Geophys Res Lett, 22, 1465

Ruffolo, D., Matthaeus, W. H., Chhiber, R., et al. 2020, Astrophys J, 902 (American Astronomical Society), 94

Schwadron, N. A., & McComas, D. J. 2021, Astrophys J, 909 (American Astronomical Society), 95

Smith, C. W., Mullan, D. J., Ness, N. F., Skoug, R. M., & Steinberg, J. 2001, J Geophys Res Space Phys, 106, 18625

Stansby, D. 2021, Res Notes AAS, 5 (American Astronomical Society), 189

Tessein, J. A., Matthaeus, W. H., Wan, M., et al. 2013, Astrophys J Lett, 776, L8

Usmanov, A. V., Goldstein, M. L., Ogilvie, K. W., Farrell, W. M., & Lawrence, G. R. 2005, J Geophys Res Space Phys, 110, https://agupubs.onlinelibrary.wiley.com/doi/abs/10.1029/2004JA010699

Weber, E. J., & Davis, Jr., Leverett. 1967, Astrophys J, 148, 217

Wit, T. D. de, Krasnoselskikh, V. V., Bale, S. D., et al. 2020, Astrophys J Suppl Ser, 246 (American Astronomical Society), 39